\documentclass[preprint]{aastex}

\tighten
\input psfig

\def\kms{\,{\rm km\, s^{-1}} }

\begin{document}

\title{Why is the Fraction of Four-Image Radio Lens Systems So High?}
\author{David Rusin, Max Tegmark} 
\affil{Department of Physics and Astronomy,
University of Pennsylvania, 209 S. 33rd St., Philadelphia, PA, 19104-6396}

\begin{abstract}

We investigate the frequency of two- and four-image gravitational lens systems
in the Jodrell-VLA Astrometric Survey (JVAS) and Cosmic Lens All-Sky Survey
(CLASS), and the possible implications for dark matter halo properties.  A
simple lensing statistics model, which describes lens galaxies as singular
isothermal ellipsoids with a projected axis ratio distribution derived from
the surface brightness ellipticities of early-type galaxies in the Coma
cluster, is ruled out at the 98\% level since it predicts too few four-image
lenses (quads).  We consider a range of factors that may be increasing the
frequency of radio quads, including external shear fields, mass distributions
flatter than the light, shallow lensing mass profiles, finite core radii,
satellite galaxies, and alterations to the luminosity function for faint
flat-spectrum radio sources.  We find that none of these mechanisms provide a
compelling solution to the quad problem on their own while remaining
consistent with other observational constraints.

\end{abstract}


%\clearpage

\section{Introduction}

Most arcsecond-scale gravitational lens systems consist of two or four images,
consistent with a population of galaxies described by smooth and
centrally-concentrated mass distributions (e.g.\ Blandford \& Kochanek 1987).
The number of lensed images provides important information about galaxies,
their inner mass profiles, and potentially their environments.  For example,
observed image multiplicities have placed powerful bounds on the inner few
$h^{-1}$ kpc of lensing galaxy mass distributions, based on the absence of
detectable central images (Rusin \& Ma 2001; Norbury et al.\ 2001).  These
studies have demonstrated that the lensing mass is unlikely to fall off much
more slowly than isothermal ($\rho \propto r^{-2}$). The relative frequency of
two- and four-image systems is another key quantity, and offers a
cosmology-independent check of the various assumptions made in both
statistical and modeling studies of gravitational lensing.

Radially symmetric mass distributions produce two or three images, depending
on the steepness of the profile and the existence of a core.  Departures from
spherical symmetry, either in the form of intrinsic ellipticity or external
shear, allow for the creation of four- or five-image systems (Blandford \&
Kochanek 1987; Kormann, Schneider \& Bartelmann 1994).  The ``standard model''
of gravitational lensing statistics, which describes lensing galaxies as
nearly singular isothermal deflectors with moderate ellipticities residing in
rather small external shear fields, predicts that two-image lens systems
should dominate four-image systems in the number counts. However, previous
studies (King \& Browne 1996; Kochanek 1996b) suggested that the fraction of
four-image lens systems found in radio lens searches appears to be
surprisingly high relative to what would be expected if the typical mass
flattening of elliptical galaxies were similar to that of the light. In the
most detailed analysis of this problem to date, Keeton, Kochanek \& Seljak
(1997; hereafter KKS) showed that the image multiplicities and mass models for
a small sample of radio and optical lenses were not statistically inconsistent
with an ellipticity distribution derived from the surface brightness axial
ratios of elliptical (E) and S0 galaxies in the Coma cluster. Over the last
several years, however, a large number of new gravitational lens systems have
been discovered by the Cosmic Lens All-Sky Survey (CLASS; Myers et al.\ 1995,
1999, 2001). The fraction of quads is quite high in the expanded sample, again
suggesting the possibility of a ``quad problem.''

In this paper we re-examine the issue of image multiplicity using the largest
homogeneously selected sample of radio-loud gravitational lenses, and
investigate a wide range of factors that could be influencing the distribution
of two- and four-image lens systems. \S 2 describes the lens sample and parent
source population. \S 3 calculates the frequency of four-image systems
predicted by singular isothermal ellipsoids (SIEs) following the ellipticity
distribution of E/S0 galaxies in Coma, and confirms that there is a quad
problem: the model fails to account for the large fraction of four-image
lenses in the sample. In \S 4 we discuss potential solutions to the problem,
quantifying the effect of flattened mass distributions, external shear fields,
shallow lensing mass profiles, finite core radii, satellite galaxies, and
alterations to the luminosity function for faint flat-spectrum radio
sources. \S 5 briefly summarizes our results and discusses future work that
might shed more light on this interesting issue.

\section{Radio Lens Sample and Source Population}

Lensing statistics calculations require a homogeneous sample with known
selection criteria and a well-understood parent source population. The largest
uniformly selected sample of gravitational lens systems belongs to the
combined Jodrell-VLA Astrometric Survey (JVAS; Patnaik et al.\ 1992a; Browne
et al.\ 1998; Wilkinson et al.\ 1998; King et al.\ 1999) and Cosmic Lens
All-Sky Survey (CLASS; Myers et al.\ 1995, 1999, 2001). Using the Very Large
Array (VLA), these surveys have imaged $\simeq 15000$ flat-spectrum ($\alpha
\geq -0.5$, $S \propto \nu^{\alpha}$) radio sources down to a 5 GHz flux
density of 30 mJy, with the aim of discovering new gravitational lens systems
for a variety of cosmological and astrophysical studies. At least 18 new
lenses have been discovered, and the final set of follow-up observations is
nearly complete.

\subsection{The Lenses}

The list of JVAS/CLASS lenses is given in Table 1. This includes seven quads,
seven doubles and the six-image system B1359+154 (Rusin et al.\ 2000,
2001b). B1933+503 (Sykes et al.\ 1998) contains ten total images of a
multi-component source (four images each of the flat-spectrum core and one
steep-spectrum lobe, plus two images of the other lobe). B1938+666 (King et
al.\ 1997) contains six images of a core-jet source, with the core lensed into
four images. B2114+022 (Augusto et al.\ 2001) features four compact radio
components, but two of these could be associated with the lensing mass. In
addition, recent CLASS radio follow-up observations have identified a strong
two-image lens candidate that may be the final lens discovered in the
JVAS/CLASS survey.

To ease the analysis of the radio data, we make cuts on the above sample. Four
systems appear to be lensed by more than one primary galaxy: B1127+385
(Koopmans et al.\ 1999), B1359+154, B1608+656 (Koopmans \& Fassnacht 1999) and
B2114+022. We exclude these from our sample because compound mass
distributions consisting of multiple galaxies of comparable masses complicate
the expected image multiplicities (Kochanek \& Apostolakis 1988; Rusin et al.\
2001b). In addition, two systems are lensed by spirals rather than
ellipticals: B0218+357 (Biggs et al.\ 1999) and B1600+434 (Koopmans, de Bruyn
\& Jackson 1998). These should also be excluded as we limit our multiplicity
models to those consisting of early-type galaxies.\footnote{While some of the
newer lenses in our sample do not have good morphologically- or
spectroscopically-determined galaxy identifications at this time, the vast
majority of known lenses are ellipticals (Kochanek et al.\ 2000). We will
therefore assume that all of the galaxies not explicitly known to be spirals
are early-type.}  Furthermore, while both B1933+503 and B1938+666 feature
quadruply-imaged compact cores that contain more than 30 mJy of flux density
at 5 GHz, the latter system was identified based on the observation of
extended radio emission. Such lenses are likely to have different selection
functions and magnification biases, so we remove B1938+666 from the
sample. B1933+503 was identified from its quadruply-imaged core, so we count
it as an additional quad. Finally, we assume that the one remaining strong
lens candidate is a genuine two-image lens, and include it in this study. The
sample we employ for our analysis therefore consists of seven quads and five
doubles.

Several of the lenses we include in the sample are still more complicated than
isolated galaxies. External shear can play an important role in determining
the lensing properties of individual systems, and the potentials of at least
four lenses have strong and identified shear contributors: B1422+231 (Hogg \&
Blandford 1994), B1600+434, B2045+265 (Fassnacht et al.\ 1999) and B2319+051
(Rusin et al.\ 2001a). Three additional lens systems -- MG0414+054 (Schechter
\& Moore 1993), B1152+199 (Myers et al.\ 1999; Rusin et al.\ 2001c), and
possibly B1030+074 (Xanthopoulos et al.\ 1998) -- are dominated by large
ellipticals, but also have faint satellite galaxies near the Einstein
radius. In the case of MG0414+054, the companion plays an important role in
the mass model, and is a necessary component for fitting the large number of
constraints (Ros et al.\ 2000). We will explore the effect of large external
shears and faint satellite galaxies in \S 4.

\subsection{Source Luminosity Function}

The image multiplicity distribution depends not only on the properties of lens
galaxies, but also on the luminosity function of the parent source population
(e.g.\ Turner, Ostriker \& Gott 1984; Kochanek 1996b).\footnote{Unlike the
total lensing optical depth, image multiplicities do not depend on cosmology
or the redshifts of lenses and sources. These quantities merely scale the
caustic structure of our simplistic deflectors in a self-similar way, thereby
preserving the relative frequencies of different image geometries.}  Lensed
sources are always magnified, and are therefore transferred to higher flux
density bins. Consequently, some sources appear in a flux-limited sample only
because they are lensed. This magnification bias can significantly increase
the probability of observing a gravitational lens system, beyond the intrinsic
optical depth. Bias alters the lensing probability of sources in a bin of
total flux density ($S$) by the factor (Turner, Ostriker and Gott 1984; Maoz
\& Rix 1993)
\begin{eqnarray}
B(S, z) = {\int \phi \left({S \over \mu}, z\right) P(\mu) {1 \over \mu} d\mu
\over \phi(S, z)}
\end{eqnarray}
where $\phi(S, z)$ is the source luminosity function and $P(\mu)$ is the
distribution of total magnifications ($\mu = \sum_i \mu_i$, where the
magnification of the $i$th image is $\mu_i$) produced by a deflector for a
certain class of imaging. Note that for a source population following a
power-law luminosity function $\phi(S, z) = dn/dS \propto S^{-\eta}$, the bias
reduces to $B = <\mu^{\eta-1}>$, independent of flux density.  Therefore,
source populations with steeper number-flux relationships and deflector models
with larger average magnifications yield higher bias factors.  The steepness
of the number-flux relation is particularly important to a discussion of image
multiplicities. Because four-image lens systems are, on average, more
magnified than doubles, bias significantly enhances the fraction of quads in a
sample relative to the ratio of areas within the two- and four-image caustic
regions of a given deflector model.

The CLASS sample is dominated by sources with 5 GHz flux densities of $30 \leq
S_5 \leq 200$ mJy, while the brighter JVAS sample is limited to $S_{5} > 200$
mJy. Due to the small number of JVAS/CLASS sources with measured redshifts
(Marlow et al.\ 2000), there is little evidence for or against any redshift
dependence of the number-flux relation, so we will follow Helbig et al.\
(1999) and assume that this is independent of redshift. We determine the
approximate number-flux relation of flat-spectrum sources by fitting the 30 --
200 mJy sample to a power-law. We group the sources into bins of size $5$ mJy,
and evaluate the fit statistic $\chi^2 = \sum_i^{N_B} (N_i - n_i)^2 / n_i$,
where $N_B = 34$ is the number of bins, $N_i$ is the number of sources in the
$i$th bin and $n_i$ is the number predicted by the power-law model. A plot of
$\Delta \chi^2 / N_B$, normalized so that the best-fit model has $\chi^2 / N_B
= 0$, is plotted in Fig.~1. We find that $\eta = 2.07 \pm 0.11$ ($\Delta
\chi^2 / N_B \leq 1$), and will assume $\eta = 2.1$ throughout this paper
unless specified. Note that the flat-spectrum CLASS sources have a steeper
number-flux relation than predicted by the Dunlop \& Peacock (1990) luminosity
function ($\eta \simeq 1.8$ for the faint sources relevant to this
sample). Because both KKS and King \& Browne (1996) assumed this shallower
slope in their calculations, it is likely that they underestimated the
expected fraction of quads in radio lens samples.

Magnification bias means that many CLASS lens systems are lensed from a source
population that is an order of magnitude fainter than the nominal flux density
cutoff of the survey. A complete statistical analysis of CLASS therefore
requires an understanding of the flat-spectrum source population in the $3-30$
mJy range. Because very little is known about such faint radio sources, the
direct measurement of the flat-spectrum radio luminosity function down to a
few mJy is one of the key projects currently being undertaken by the CLASS
Consortium for European Research on Extragalactic Surveys (CERES) team (McKean
et al.\ 2001, in preparation). For this paper, however, we will make the
simplest possible assumption: that the slope of the number-flux relation
remains unchanged as we move to fainter flux density levels.  Is such an
assumption compatible with our observations? A simple quantitative check
involves comparing the respective lensing rates of JVAS and CLASS. Because the
surveys are lensed by the same galaxy population -- and obviously probe the
same cosmology! -- the relative lensing rates depend only on the redshift
distribution of the samples and the number-flux relations of the parent source
populations. The redshift distribution of flat-spectrum sources in the $25-50$
mJy range has been investigated (Marlow et al.\ 2000) and appears to be quite
similar to that of much brighter sources, though at a lower level of
completeness. In fact, the mean redshift of flat-spectrum radio sources
experiences remarkably little evolution over more than two orders of magnitude
in flux density (Falco, Kochanek \& Mu\~noz 1998; Marlow et al.\
2000). Differences in the observed lensing rates should therefore be largely
attributed to differences in the respective number-flux relations. However,
preliminary results indicate that the lensing rate in the combined JVAS/CLASS
sample ($\simeq 1/600$) is compatible with that of the JVAS sample alone. We
therefore believe that an assumption that the number-flux relation remains
unchanged at low flux density levels is a reasonable and conservative one at
this time.

\section{A Simple Model for Image Multiplicities}

In this section we test a simple model for predicting the frequency of
different image geometries in radio lens surveys. The model is based on the
following three assumptions: 1) elliptical lens galaxies can be described as
singular isothermal ellipsoids, 2) the mass ellipticities of the E/S0 galaxy
population at $z\simeq 0.5$ are similar to the light ellipticities of nearby
cluster-bound early-type galaxies, and 3) the typical external shear fields in
which lensing galaxies sit are negligible for statistical lensing
considerations. We discuss the motivation for these assumptions, then
determine whether the JVAS/CLASS multiplicity data is compatible with the
model.

The first assumption concerns the mass distribution in lensing
galaxies. Virtually all evidence from gravitational lensing, and in particular
from the modeling of highly-constrained lens systems (Kochanek 1995; Grogin \&
Narayan 1996; Romanowsky \& Kochanek 1999; Cohn et al.\ 2000), suggests that
the lensing mass distribution of elliptical galaxies is close to
isothermal. The isothermal paradigm for early-type galaxies is strongly
supported by investigations of X-ray halos (Fabbiano 1989) and stellar
dynamics (Rix et al.\ 1997). Furthermore, the absence of central images in
deep radio maps of gravitational lens systems argues that the core radii must
be quite small (e.g.\ Wallington \& Narayan 1993; Kochanek 1996a), with the
best constrained lenses requiring cores of less than a few tens of parsecs
(Norbury et al.\ 2001).  All lensing data are currently consistent with zero
core radius.

We therefore approximate the lensing mass distribution in galaxies as a
singular isothermal ellipsoid (SIE; Kormann, Schneider \& Bartelmann 1994).
For a galaxy with central velocity dispersion $\sigma_v$, the scaled surface
mass density is
\begin{eqnarray}
\kappa(\theta_1, \theta_2) = {b_{E} \over 2
(\theta_1^2 + f^2 \theta_2^2)^{1/2} }
\end{eqnarray}
where $b_E = b_{SIS} \omega(f)$, $b_{SIS} = 4 \pi (D_{ds}/D_s) (\sigma_v/c)^2$
(with $D_{ds}$ and $D_s$ the angular diameter distances from the lens to the
source and from the observer to the source, respectively), $f$ is the axial
ratio and $\omega(f)$ is an additional ellipticity-dependent normalization
factor. When comparing SIEs with different flattenings, the normalization
should be chosen so that the circular velocity (or central line-of-sight
stellar velocity dispersion) is independent of axial ratio (KKS). Following
the arguments and calculations of KKS and Keeton \& Kochanek (1998), we set
$\omega(f) = (1-f^2)^{1/2} / \tan^{-1} (1/f^2 - 1)^{1/2}$. For simplicity all
galaxies are assumed to be viewed edge-on. The lensing cross-section $\sigma
\propto \omega(f)^2$, and increases as the mass distribution is made flatter
when the above normalization is employed.

Fig.\ 2 concisely illustrates the effect of ellipticity on the lensing
properties of singular isothermal deflectors. Spherical models have a single
radial caustic curve on the source plane, inside of which two-image lens
systems are formed. As the model is made more elliptical, a tangential
(diamond) caustic emerges from the core and increases in size. Sources within
both caustics produce four-image lens systems. Deflectors with small
ellipticities preferentially produce two-image lens systems, but the fraction
of quads increases rapidly as the mass distribution is flattened. The most
elliptical deflectors are dominated by naked-cusp geometries, in which three
bright images are formed on the same side of the lens with no
counter-images. In this regime the number of quads is diluted, as some sources
within the tangential caustic are now lensed into cusp configurations.

We tabulated the lensing properties of the SIE for $0.1 \leq f \leq 1.0$ via
Monte Carlo simulations. For each $f$, sources were placed randomly behind the
mass distribution and the lens equation was numerically inverted using a
modified downhill simplex optimization routine (Press et al.\ 1992) to solve
for all image positions and magnifications. Our code makes use of the analytic
deflection angles and shear vectors available for isothermal mass
distributions (Kormann, Schneider \& Bartelmann 1994). The cross-sections
($\sigma$) and magnification biases ($B$) were then computed from this data
for each class of imaging. The resolution of the simulations ensured that the
relevant quantities were recovered to $< 1\%$. We excluded from these
calculations all lens systems with a magnification ratio between the brightest
two images of greater than 15:1, in accordance with the flux ratio selection
function appropriate for radio lens searches using the VLA. Because quads
usually contain three relatively bright images while doubles often have one of
their images significantly demagnified, the flux ratio cut rejects more
two-image lens systems and will therefore slightly increase the fraction of
quads. The relative lensing probabilities for doubles, quads, and naked-cusp
configurations are shown in Fig.~3 as a function of axial ratio $f$.

Second, we make the joint assumption that a) the projected mass axial ratios
of early-type galaxies are similar to those of the light, and b) the
population of galaxies at lensing redshifts ($z \simeq 0.5$) has similar
ellipticities to the nearby cluster-bound population. Because the lensing mass
of galaxies contains a substantial fraction of dark matter (e.g.\ Kochanek
1995), it is by no means necessary that the mass and light distributions have
the same ellipticity. Investigations of polar-ring galaxies (Sackett et al.\
1994; Combes \& Arnaboldi 1995) and the X-ray halos of ellipticals (Buote \&
Canizares 1994, 1996) have suggested that dark matter halos may be quite flat
($0.3 \leq f \leq 0.6$). This is in some conflict with recent N-body
simulations incorporating gas dynamics (Cole \& Lacey 1996), which indicate
that CDM halos may be much rounder than previously suggested by Dubinski
(1994). Both gravitational lens modeling and the statistics of image
multiplicities are potentially powerful probes of the flattening of lensing
mass distributions.

There is little evidence that distant early-type galaxies represent a
different population than those found in nearby clusters. First, photometric
studies by the CASTLES collaboration (Kochanek et al.\ 2000) demonstrate that
lensing galaxies fall on the familiar fundamental plane. Second, the typical
surface brightness axial ratios of lensing galaxies, which are found mostly in
the field or poor groups (Keeton, Christlein \& Zabludoff 2000), appear to be
similarly distributed to those in the Coma cluster. Based on the measurements
of Jorgensen \& Franx (1994), KKS find that the Coma E/S0 galaxies are well
described by a gaussian ellipticity distribution $dn(\epsilon)/d\epsilon
\propto \exp [-(\epsilon - \epsilon_o)^2/2 \Delta \epsilon^2]$, where
$\epsilon = (1-f^2)/(1+f^2)$, $\epsilon_o = 0.26$ and $\Delta \epsilon =
0.33$. This function is plotted in Fig.\ 4, and we will refer to the
distribution as the ``Coma ellipticity model.'' We compare the Coma model with
the surface brightness ellipticities of lensing galaxies listed in Keeton,
Kochanek \& Falco (1998) via a Kolmogorov-Smirnov test, and find that the two
distributions are 76\% compatible. Therefore, it is likely that the light
ellipticities of lensing galaxies are drawn from a distribution similar to
that of the the cluster-bound E/S0 galaxies in Coma.

The third assumption is that the external shear that typical lensing galaxies
experience should be small. Using a shear model based on the galaxy-galaxy
correlation function, which should be a good approximation if most lens
galaxies are not embedded in groups or clusters (this is compatible with
observations), KKS argue that galaxies tend to experience small external
shears ($\gamma \simeq 0.03$), with only a small fraction of lenses ($\simeq
5\%$) expected to have shear perturbations of $\gamma > 0.10$. When external
shears are small, the most important source of shear is internal, due to the
ellipticity of the primary lensing galaxy itself. Furthermore, KKS claim that
the influence of external shear on lensing cross-sections would be comparable
to that of ellipticity only if the perturbations were about an order of
magnitude larger than what is expected from their simple model. Optical
observations and mass modeling of gravitational lens systems, however, suggest
that the effect of nearby galaxies may be much higher than predicted by the
correlation function model. Four lenses in our sample have significant and
well-identified external shear contributors, which are required to provide
reasonable fits to the image data. We will reopen the question of external
shear fields in the following section, but for now will retain the assertion
that shear fields are small enough to be insignificant to lensing statistics.

We now compute the fraction of four-image lens systems produced by a
population of SIEs following the Coma ellipticity model. To accomplish this we
integrate over the model to find the expected lensing optical depth (up to a
constant) for each class of imaging: 
\begin{eqnarray}
\tau_k = \int B_k(f) \sigma_k(f) {dn(f) \over df} df 
\end{eqnarray}
where $\sigma_k(f)$ is the unbiased cross-section, $B_k(f)$ is the
magnification bias factor including the flux ratio cut, and $dn(f)/df$ is the
normalized distribution of axial ratios derived from the Coma ellipticities
(Fig.~4). We cut off the distribution at $f=0.1$ for all calculations.  The
fraction of lens systems of a given class is then simply $p_k = \tau_k / \sum
\tau_j$. We find that the Coma model produces a quad fraction of $p_Q =
0.273$. This value changes to $p_Q = 0.244$ if $\eta = 2.0$ and $p_Q = 0.306$
if $\eta = 2.2$.  The predicted quad fraction is much lower that the observed
fraction of $\simeq 60\%$. (If the Dunlop \& Peacock (1990) value of $\eta =
1.8$ were assumed, $p_Q = 0.193$, so the discrepancy would be worse.)

Since very flattened galaxies do not contribute significantly to the
statistics of the SIE/Coma model, the probability of producing cusp-imaged
triples is negligible. Therefore, the likelihood that a sample of $N$ radio
lenses includes $N_Q$ four-image systems is governed by binomial statistics:
\begin{eqnarray}
L(N, N_Q) = {N! \over N_Q ! (N-N_Q)!} p_Q^{N_Q} (1-p_Q)^{N-N_Q}
\end{eqnarray}
Based on the above calculations, the probability for producing a set of $N =
12$ lenses with fewer than $N_Q$ quads is plotted in Fig.~5. The probability
of our sample containing $N_Q < 7$ quads is 97.7\%.  In other words, the basic
model for image multiplicities is ruled out at about the 98\% level by the
high fraction of four-image lenses in the JVAS/CLASS survey.

\section{Why So Many Quads?}

We have demonstrated that there exists a statistically significant overdensity
of four-image lens systems in the JVAS/CLASS radio sample, despite that fact
that the true number-flux relation of flat-spectrum radio sources leads to a
higher quad fraction than predicted in previous analyses.  We now consider a
wide range of factors that may be increasing the frequency of quads, including
flattened mass distributions, external shear, galaxy core radii,
non-isothermal mass profiles, satellite galaxies, and a source luminosity
function break below the CLASS survey cutoff. In each case we quantify the
effect on the distribution of two- and four-image lens systems, and evaluate
whether the corrections can account for the large number of quads found in the
JVAS/CLASS survey while retaining consistency with other observations.

\subsection{Very Flattened Dark Matter Halos}

The most straightforward way of producing more quads is to make the mass
distributions of lensing galaxies flatter than the light distributions of
typical ellipticals. How flat must the characteristic galaxy mass distribution
be to reproduce the JVAS/CLASS results? From Fig.\ 3 it is clear that the
lensing data will bound the range of preferred SIE axial ratios on both
ends. Large axial ratios will be excluded because they cannot produce a
sufficient fraction of quads. Small axial ratios will be excluded because they
are dominated by naked-cusp configurations, which are not observed in our
sample (or anywhere).

To analyze the probability of a given projected axial ratio producing a sample
of $N$ lenses with $N_Q$ quads, $N_D$ doubles and zero naked-cusps, we employ
the trinomial distribution. Therefore, if the fraction of doubles produced is
$p_D(f)$, while that of quads is $p_Q(f)$, the relative likelihood of the model
is
\begin{eqnarray}
L(f) = {N! \over N_Q ! N_D !} p_Q(f)^{N_Q} p_D(f)^{N_D}
\end{eqnarray}
for the observed case where naked-cusp systems are absent. The results are
plotted as part of Fig.\ 7d, normalized to the peak likelihood of unity. Using
the SIE model, the JVAS/CLASS data require a typical mass axial ratio of $f =
0.40^{+0.26}_{-0.14}$(95\% confidence).

While this solution has not been definitively ruled out, there are two
potential inconsistencies with such flat mass distributions. First, analyses
of mass-to-light ratios suggest that the lensing mass of galaxies contains
approximately equal fractions of dark and luminous matter (e.g.\ Kochanek
1995; Jackson et al.\ 1998b). Because it is reasonable to assume that the mass
distribution associated with the luminous component should have the same
ellipticity as the light, extremely flattened dark matter distributions would
be required to produce a total projected mass distribution with $f\simeq
0.4$. No currently popular models of dark matter predict halos this flat
(e.g.\ Cole \& Lacey 1996; Dav\'e et al.\ 2000). Second, if the mass
distributions of galaxies are flatter than the light, there should be a
systematic discrepancy between the mass axial ratios required by models of
individual lens systems and the surface brightnesses of their lensing
galaxies. Keeton, Kochanek \& Falco (1998) present a variety of models and
photometric data for more than a dozen systems lensed by early-type galaxies,
and in most cases the mass and light ellipticities are consistent. An
extensive and systematic modeling study of all gravitational lens systems
should provide important evidence for or against the mass flattening
hypothesis.

\subsection{External Shear Fields}

External shear can significantly affect the radial asymmetry of lensing
potentials, and is an important component in models of highly-constrained lens
systems (Hogg \& Blandford 1994; Ros et al.\ 2000; Rusin et al.\ 2001a). KKS
demonstrated that in many cases, lenses can be modeled equally well using
either an SIE or a singular isothermal spheroid (SIS) plus shear. Furthermore,
the models lead to similar quad-to-double ratios if $\epsilon \simeq 3
\gamma$, where $\gamma$ is the magnitude of the external shear
field. Intrinsically elliptical deflectors in external shear fields are more
complicated, as one must consider the relative orientation angle ($\phi$)
between the shear field and the galaxy major axis. When the internal and
external shear axes are well-aligned ($\phi \simeq 0^{\circ}$), the potential
has a much larger effective ellipticity and more quads are formed relative to
the pure SIE case. Misalignment has the opposite effect, and produces a
``rounder'' potential in which fewer quads can form. Therefore, one expects
much of the effect of external shear to cancel out if the field is averaged
over relative orientations, particularly for SIEs in which the internal shear
term dominates ($\epsilon > 3 \gamma$).

To evaluate the effect of shear on lensing statistics, we investigated the
lensing properties of SIE deflectors in external shear fields of strength
$\gamma = 0.05$ and $0.10$.  We repeated the trials for different orientation
angles $\phi = 0^{\circ}$ to $90^{\circ}$, in increments of $5^{\circ}$. Due
to reflection symmetry, the analysis can be limited to a single quadrant. We
then integrated the two- and four-image cross-sections over $\phi$, assuming
that the external shear is randomly oriented relative to the position angle of
the galaxy major axis. The resulting frequency of quads, as a function of
axial ratio $f$, is plotted in Fig.~6. As anticipated, the effect of the
random shear fields is limited to the regime of small ellipticity ($f \ga
0.7$). Integrating over the Coma ellipticity model, we find that $p_Q = 0.291$
for $\gamma = 0.05$ and $p_Q = 0.328$ for $\gamma = 0.10$. In these cases, the
probabilities of producing a 12-lens sample with fewer than 7 quads are 96.7\%
and 93.9\%, respectively.

Estimates of the typical external shear fields experienced by lenses, both
from the galaxy-galaxy correlation function (KKS) and large-scale structure
(``cosmic shear'') studies (Barkana 1996), predict rather small
shears. According to the KKS analysis, only $\simeq 5\%$ of galaxies should
experience shears in excess of $\gamma = 0.10$. Assuming an isothermal model,
$\gamma = 0.10$ is approximately the shear produced by an $L_*$ elliptical
galaxy at a distance of $3\farcs 5$ from the primary deflector, or a $400\kms$
group/cluster at $11''$. Because companions of this size have not been
recognized for many lens galaxies, it is questionable whether the effects of
shear alone can account for the high frequency of quads.

\subsection{Core Radii}

One can also increase the fraction of quads by decreasing the number of
doubles produced by the deflector population. This may be achieved if the
inner mass profile of lensing galaxies is shallower than isothermal, either by
introducing a core radius or decreasing the power-law slope of the profile
(Blandford \& Kochanek 1987; Kassiola \& Kovner 1993; Wallington \& Narayan
1993; Kormann, Schneider \& Bartelmann 1994). We will discuss cores first.
Fig.\ 1 concisely illustrates the effect of core radii on lensing
statistics. A core decreases the deflection angle near the center of the lens
and thereby shrinks the size of the radial caustic, inside of which doubles
form. However, the size of the tangential caustic depends only on the
ellipticity of the potential, and is virtually undisturbed as the profile is
modified. Consequently, for rounder deflectors, a core will tend to decrease
the number of doubles while leaving the number of quads unmodified. For
flatter deflectors, the smaller radial caustic attributed to the core implies
that naked-cusp configurations become prominent at larger axial
ratios. Because a fraction of the area inside the tangential caustic will form
cusp systems, large cores dilute the quad fraction for high ellipticity
deflectors.

Before continuing, we should note that introducing core radii to the
isothermal model or decreasing the slope of the mass profile produces an
additional lensed image near the center of the lensing galaxy (Rusin \& Ma
2001; Norbury et al.\ 2001). The slower the mass distribution falls with
radius, the brighter these images should appear. In this subsection, systems
that we refer to as ``doubles'' and ``quads'' will actually contain an
additional central component. Naked-cusp configurations still do not feature
any counter-images.  The absence of central images places powerful bounds on
how far one can safely stray from a singular isothermal mass model, and we
return to these priors below.

We consider the non-singular isothermal ellipsoid (NIE), which is described by
the scaled surface mass density:
\begin{eqnarray}
\kappa(\theta_1, \theta_2) =
{b_E \over 2 (\theta_1^2 + f^2 \theta_2^2 + \theta_c^2)^{1/2} }
\end{eqnarray}
where $\theta_c$ is the angular core radius. The lensing properties of the NIE
are determined by the axial ratio $f$ and the ratio between the core radius
and the angular Einstein radius, $\theta_c / b_E$. Again we performed a series
of Monte Carlo simulations of the NIE over a grid of parameters ($0.1 \leq f
\leq 1.0$, $0 \leq \theta_c / b_E \leq 0.1$), calculating the cross-sections
and magnification bias for each model. Angular core radii $\theta_c / b_E \ga
0.1$ correspond to linear cores of several hundred parsecs at lensing
redshifts, and are strongly ruled out by virtually all lensing data (Norbury
et al.\ 2001). If $\theta_c / b_E \ga 0.5$, the deflector cannot produce
multiple images (Kormann, Schneider \& Bartelmann 1994). The magnification of
the central image is included in the determination of the bias, and we make no
cuts on its brightness.

The frequency of quads as a function of $f$ is plotted for $\theta_c / b_E =
0, 0.025, 0.050, 0.075$ and $0.100$ in Fig.~7a. Adding large cores
significantly increases the quad fraction for small to moderate
ellipticities. For example, an $f=0.7$ deflector produces only 22\% quads for
the singular case, but this increases to 27\% if $\theta_c / b_E = 0.025$ and
43\% if $\theta_c / b_E = 0.100$. However, the quad fraction is decreased at
the high ellipticity end, as a large fraction of sources inside the tangential
caustic form cusp-imaged triples (Fig.~7b). The total fractions of quads and
cusps, integrated over the Coma ellipticity model, are plotted in Fig.~7c as a
function of $\theta_c / b_E$. Incredibly, the fraction of quads increases very
little with increasing core radii. The effect is due to the specific
parameters of Coma model, which lead to a dilution of quads at the high
ellipticity end that almost completely cancels the increase in quads at the
low ellipticity end. Therefore, if one assumes that the mass axial ratios of
lensing galaxies are accurately described by the Coma model, any core radius
does little to affect the overall fraction of quads produced. Finally, the
typical axial ratios required to fit the lensing data, for various core radii,
are calculated using (5) and plotted in Fig.\ 7d. Larger cores mean that the
observed quad fraction can be reproduced at smaller characteristic
ellipticities.

The lack of central images in deep maps of radio-loud gravitational lens
systems implies that the core radii of lensing galaxies are unlikely to be
very large. Norbury et al.\ (2001) use the absence of third images to
constrain the size of the core in isothermal deflectors, and find that a
number of lenses must have core radii smaller than a few tens of parsecs. For
an $L*$ elliptical at $z = 0.5$, $\theta_c / b_E = 0.025$ (the smallest core
we considered in the above simulations) corresponds to a linear core radius of
$\simeq 150$ pc assuming a flat $\Omega_m = 0.3$ cosmology with $h =
0.65$. Therefore, the extremely small cores allowed by lensing are unlikely to
have any significant effect on the predicted image multiplicities.

\subsection{Shallow Mass Profiles}

Decreasing the power-law slope of the radial mass profile has a similar effect
on the caustic structure as adding a core radius to an isothermal mass
distribution (e.g. Rusin \& Ma 2001). We consider singular mass distributions
described by a power-law surface density of the form
\begin{eqnarray}
\kappa(\theta_1,\theta_2) = {b_E \over 2 (\theta_1^2 + f^2 \theta_2^2
)^{\beta/2 }}
\end{eqnarray}
where $\beta$ is the slope of the mass profile. For the isothermal case,
$\beta = 1$. We performed Monte Carlo analyses of the lensing properties for
the singular power-law ellipsoid (SPLE) as a function of $f$ and $\beta$. Our
calculations make use of the rapidly converging series solutions for the
deflection angles and magnification matrices of power-law mass distributions
derived by Barkana (1998) and implemented in the ``FASTELL'' software
package. The results are plotted in Fig.\ 8. As was the case when adding cores
to isothermal deflectors, decreasing the slope of the mass profile
significantly increases the quad fraction for small ellipticities and dilutes
it for large ones (Fig.~8a). Naked-cusp systems again become prominent at
larger axial ratios as the mass distribution is made shallower
(Fig.~8b). Assuming a Coma ellipticity distribution, the integrated fractions
of quads and cusps are plotted in Fig.\ 8c, as a function $\beta$. Once again,
the Coma model cannot produce arbitrarily many quads. The typical axial ratios
required to fit the lensing data, for various profiles, are calculated using
(5) and plotted in Fig.\ 8d. It is interesting to note from Fig.\ 7a and 8a
that regardless of axial ratio, core radius or mass profile, the largest
fraction of quads any deflector can produce is $\simeq 60\%$. This was first
pointed out by KKS for the singular isothermal case. Because the observed
fraction of quads is nearly equal to this maximal value, the typical
properties of galaxy mass distributions would have to be very finely tuned to
fit the data. It is therefore not surprising that when integrating over a
distribution of deflectors, we consistently fall far short of the preferred
quad fraction. This is perhaps the most unsettling aspect of the quad problem.

The lack of detectable central images suggests that the mass distribution in
lensing galaxies is unlikely to be much shallower than isothermal. Rusin \& Ma
(2001) demonstrate that the probability of finding no central images among the
CLASS doubles falls off rapidly as the power-law mass profile is decreased,
thereby strongly excluding characteristic profiles with $\beta <
0.8$. Moreover, direct modeling of highly-constrained lens systems (e.g.\
Kochanek 1995; Cohn et al.\ 2000), modern stellar dynamics (Rix et al.\ 1997)
and X-ray halos (Fabbiano 1989) all suggest that the lensing mass distribution
of early-type galaxies is close to isothermal. Therefore, shallow mass
profiles are unlikely to be to blame for the large fraction of radio quads.

\subsection{A Broken Faint-end Luminosity Function}

Steeper luminosity functions lead to more four-image lens systems. As
described in $\S 2$, the power-law number-flux law obeyed by flat-spectrum
radio sources in the $30-200$ mJy range is tightly constrained to lie at $\eta
\simeq 2.1$. In our subsequent calculations, we made the assumption that the
radio luminosity function in the $3-30$ mJy range is identical to that of the
brighter CLASS sources, consistent with the relative lensing rates in JVAS and
CLASS. Here we briefly evaluate our assumption in greater detail, exploring
the possible effects of a faint-end luminosity function with a break below the
CLASS cutoff.

Assuming that the $30-200$ mJy source population is well-described by $\eta =
2.1$, a break below $30$ mJy would have a maximal impact on the statistics of
CLASS if it occurs right at the flux cutoff of the survey. We therefore
investigate a broken power-law number-flux relation of the form
\begin{eqnarray}
\phi(S,z) \propto \left\{ \begin{array}{l l}
	(S/S_o)^{-\eta}     & S \geq S_o \\
	(S/S_o)^{-\eta_f}   & S \leq S_o
	
	\end{array}\right.
\end{eqnarray}
where $S_o = 30$ mJy, the limiting CLASS 5 GHz flux density. A consequence of
any luminosity function more complicated than a single power-law is that the
magnification bias is no longer independent of source flux density. We
therefore consider representative cases of $S = 30, 50, 100$ and $300$ mJy for
our calculations. We fix $\eta = 2.1$ and revert back to the Coma/SIE model to
investigate the effects of a broken luminosity function on the quad fraction
and lensing rates.

The predicted fraction of quads ($p_Q$) is plotted against $\eta_f$ in Fig.\
9a. As expected, steeper faint-end power-laws produce a larger number of quads
than shallower ones, due to the increased magnification bias.  The total
lensing optical depth, relative to the unbroken case of $\eta_f = 2.1$, is
shown in Fig.\ 9b. A larger $\eta_f$ also leads to a significantly increased
lensing rate for the CLASS $30-100$ mJy sources. However, typical $300$ mJy
JVAS sources are only slightly affected by the break. One requires $\eta_f
\simeq 2.5$ for a moderately respectable quad fraction of $p_Q = 0.4$ in
CLASS, but this will lead to a lensing rate among $30-50$ mJy sources that is
more than twice that of the $300$ mJy population. Because the observed lensing
rates in JVAS and CLASS are both $\simeq 1/600$, we conclude that
modifications of the radio number-flux relation are unlikely to account for
the high fraction of quads in our sample, while at the same time retaining
consistency with the observed relative lensing rates. However, it is still
very important to measure the faint-end luminosity function directly, for the
purposes of both better investigating the quad problem and obtaining robust
constraints on the cosmological density parameters through a full analysis of
the CLASS survey statistics.

\subsection{Faint Satellite Galaxies}

Yet another possibility is that faint satellite galaxies are strongly
affecting the caustic structure of lens systems. While lenses of nearly equal
mass should not be common (Kochanek \& Apostolakis 1988), many primary lens
galaxies may have small satellites because the galaxy luminosity function
diverges at the low luminosity end. Faint companions have been observed in
several JVAS/CLASS lens systems -- MG0414+054 (Schechter \& Moore 1993),
B1152+199 (Rusin et al.\ 2001c), and possibly B1030+074 (Xanthopoulos et al.\
1998). For the highly-constrained four-image lens MG0414+054, the satellite is
an important element of the lens model (Ros et al.\ 2000).

A systematic investigation of the influence of companion galaxies on the quad
fraction would be extremely useful, but requires very computationally
expensive simulations to sufficiently cover the large parameter space. While
such a study is beyond the scope of this paper, we did perform a brief series
of Monte Carlos to get a feel for the magnitude of the statistical correction
due to companion galaxies. We modeled the primary galaxy as an $L_*$ SIE
($\sigma_v = 220 \kms$) with $f=0.7$, and the satellite as an SIS. We
considered cases with $b_{SIE} = 5 b_{SIS}$ and $b_{SIE} = 10 b_{SIS}$, and
assumed a flat $\Omega_m = 0.3$ cosmology with lens and source redshifts of $z
= 0.5$ and $z=1.5$, respectively, to convert to real units. Based on the
observed separation range of satellites, we placed the companions at $r =
0\farcs5$, $1\farcs0$ and $1\farcs5$ from the center of the primary lens. We
repeated the analysis for different relative orientation angles ($\phi$)
between the major axis of the primary galaxy and the position angle of the
satellite ($0^{\circ} \leq \phi \leq 90^{\circ}$), in increments of
$5^{\circ}$. We assumed randomly-oriented satellites and integrated over the
position angles.

Satellites can have a large effect on the lensing properties for any
particular orientation angle, greatly increasing the quad fraction when they
are well-aligned with the major axis of the primary galaxy ($\phi \simeq
0^{\circ}$), and decreasing it when they are anti-aligned ($\phi \simeq
90^{\circ}$). Interaction between the SIE and SIS caustics can also produce
more complex image configurations (see Rusin et al.\ 2001b).  However, like
shear, the orientation-integrated effect on quads is negligible. In five of
the six sets of trials we undertook, the quad fraction increased by less than
$1\%$ from the value predicted for an isolated SIE with $f = 0.7$ (22\%). For
the case of $b_{SIE} = 5 b_{SIS}$ and $r = 1\farcs0$, the relative deflector
strengths and separations produce significant overlap between the tangential
caustic of the SIE and the radial caustic of the SIS. Sources sitting within
one tangential and two radial caustics produce five observable images instead
of four, and the quad fraction is attenuated to $\simeq 17\%$. We note that
these fifth images are formed adjacent to the satellite and should be readily
detectable, so the probability of confusing such systems with true quads is
small. The superposition of caustics occurs for MG0414+054. We analyzed the
caustic structure of the published SIE+SIS+shear model (Ros et al.\ 2000) and
find that if the companion (SIS) is removed, the predicted quad fraction is
$\simeq 21\%$. Including the companion results in a significant overlap of the
SIE tangential and SIS radial caustics, decreasing the quad fraction to only
$\simeq 8\%$. Therefore, in the case of MG0414+054, the system appears to be a
quad not because of the satellite, but in spite of it.

From our preliminary analysis we find little evidence to suggest that faint
companions are significantly increasing the quad fraction produced by moderate
ellipticity galaxies. A more thorough numerical investigation of satellite
galaxies is necessary for a full accounting of their statistical impact.

\section{Discussion}

We have investigated the distribution of image multiplicities in the
JVAS/CLASS radio lens sample, and find that the large fraction of four-image
gravitational lens systems poses some very interesting astrophysical
questions. First we modeled early-type lensing galaxies at $z \simeq 0.5$ as a
population of isothermal deflectors with a projected mass ellipticity
distribution equal to that of the surface brightness ellipticities of E/S0
galaxies in the nearby Coma cluster. This is a reasonable assumption, as
lensing galaxies have similar light ellipticities to the Coma sample. Our
model predicts a quad fraction of $p_Q \simeq 27\%$, higher than the
predictions of previous analyses that used a shallower flat-spectrum source
luminosity function. However, the new prediction still falls far short of the
$\simeq 60\%$ quad fraction of the JVAS/CLASS sample, and would produce fewer
than the observed number of quads $\simeq 98\%$ of the time. The most basic
model for image multiplicities is therefore excluded at about 98\% confidence,
confirming that the old quad problem, while not statistically severe, is
stubbornly refusing to go away.

We next considered a range of potential solutions, effects that may be
increasing the fraction of radio quads. One possibility is that the mass
distributions of lensing galaxies are much flatter than the light of typical
ellipticals. We find that a mass axial ratio of $f \simeq 0.40$ is required to
fit the data. This solution may be problematic, as N-body simulations argue
against such flattened dark matter halos, and lens modeling has thus far
failed to present any convincing evidence that lensing mass distributions are
systematically flatter than the light. Another possibility is that external
shear fields are influencing the distribution of image multiplicities. If all
lensing galaxies sit in rather large shear fields of $\gamma = 0.10$, the quad
fraction increases to only $\simeq 33\%$ if the fields are randomly
oriented. While shear of this magnitude does play an important role in the
lensing potentials of several lens systems, external shear on its own is not
likely to be the source of the discrepancy. We also considered increasing the
quad fraction by decreasing the number of doubles produced by the deflector
population. This can be achieved by introducing large cores or making the mass
profiles of lensing galaxies shallower than isothermal. However, this is
unlikely to be a viable solution, as the absence of central images in deep
radio maps suggests that lensing mass distributions cannot be much shallower
than isothermal, and cannot have significant core radii. Furthermore, while
large cores or very shallow profiles can greatly enhance the number of quads
in the small ellipticity regime, the Coma model is such that nearly an equal
number of quads are lost in the high ellipticity regime. As a result, the quad
fraction cannot be made arbitrarily large using the Coma ellipticities, even
if large cores and shallow profiles were acceptable. We briefly investigated
the effect of faint satellite galaxies on deflectors with moderate
ellipticity, and find no significant changes to the quad fraction. More
investigation of this important issue is required to solidify our preliminary
conclusions. Finally, we considered modifications to the flat-spectrum source
luminosity function below the 30 mJy 5 GHz cutoff of the CLASS survey. More
quads can be produced if the number-flux relation grows steeper for low flux
densities, but this will lead to a much larger lensing rate among the fainter
CLASS sources than the brighter JVAS sources. The similar lensing rates in the
surveys strongly argue against a steepening luminosity function below 30 mJy.

Perhaps the most disturbing possibility of all is that a number of two-image
lens systems have remained unidentified in the JVAS/CLASS surveys. Could there
exist some observational bias that favors the discovery of four-image systems
over two-image systems? It is certainly true that a large fraction of quads
can be identified from the JVAS/CLASS VLA survey maps alone, while virtually
every two-image lens candidate requires an intense program of radio follow-ups
to prove or disprove the lensing hypothesis (Myers et al.\ 2001). However, the
selection of two-image candidates is very conservative, ensuring that even
extremely unpromising sources are passed through at least one level of the
high-resolution radio filter. Furthermore, deep radio imaging with the
Multi-Element Radio-Linked Interferometer Network (MERLIN) and Very Long
Baseline Array (VLBA) have proven themselves to be excellent tools for both
confirming lens candidates through the detection of correlated milliarcsecond
substructure in the radio components (e.g.\ Patnaik et al.\ 1995; Marlow et
al.\ 2001; Rusin et al.\ 2001a), and in most cases disproving the lensing
hypothesis through the observation of obvious core-jet morphologies. Any
remaining ambiguous candidates are then observed optically in the attempt to
detect a possible lensing galaxy. Therefore, while the confirmation of
two-image lens candidates can be quite demanding, we have little reason to
believe that the observational pipeline of CLASS is less than robust. In a
particularly interesting turn of events, a reanalysis of the entire data set
recently revealed a problem with the automatic candidate selection code that
had actually been biasing {\em against} the identification of small four-image
lens systems, due to the blending of closely-spaced components in the VLA
survey maps (Phillips et al.\ 2000).  A set of candidates selected with the
upgraded code has resulted in the discovery of one additional CLASS lens --
the quad B0128+437 (Phillips et al.\ 2000). We therefore believe that it is
unlikely that selection effects can account for the large radio quad-to-double
ratio.

In conclusion, none of the above-mentioned solutions to the quad problem are
particularly satisfying on its own.  Fortunately, future work is likely to
shed new light on this mystery. A detailed and systematic modeling study of
all gravitational lens systems should offer important evidence for or against
the hypothesis that mass ellipticities trace the light. These same analyses
may also help determine the typical shear fields in which gravitational lens
galaxies reside. In addition, a reanalysis of the predicted image
multiplicities may be worthwhile once the full radio luminosity function has
been directly measured by programs currently being undertaken by the
CLASS/CERES team. New well-selected lens samples, such those discovered as
part of the PMN radio lens survey (Winn et al.\ 2000) and the Sloan Digital
Sky Survey, should offer us greater statistical power in determining whether
the large fraction of quads in JVAS/CLASS is telling us something about
galaxies and their environments, or is merely a fluke.

\acknowledgements

The authors thank Roger Blandford, Phillip Helbig, David Hogg, Chuck Keeton,
Chris Kochanek, Leon Koopmans, Chung-Pei Ma, Shude Mao, Paul Schechter, David
Spergel, Matthias Steinmetz and Ed Turner for helpful comments and
discussions.  Support for this work was provided by the Zaccheus Daniel
Foundation, the University of Pennsylvania Research Foundation, NSF grant
AST00-71213 and NASA grant NAG5-9194.

\clearpage

\begin{table*}
\begin{tabular}{ l l c l l}
\hline \hline
Survey & Lens &  Images & Reference & Comments\\
\hline 
CLASS   & B0128+437 & 4	    & Phillips et al.\ 2000	& no optical imaging\\
JVAS/MG & J0414+054 & 4     & Hewitt et al.\ 1992	& faint satellite\\
CLASS   & B0712+472 & 4	    & Jackson et al.\ 1998a	& \\
JVAS    & B1422+231 & 4	    & Patnaik et al.\ 1992b	& shear \\
CLASS   & B1555+375 & 4	    & Marlow et al.\ 1999	& no optical imaging\\
CLASS   & B1608+656 & 4	    & Myers et al.\ 1995	& compound \\
CLASS   & B1933+503 & 4+4+2 & Sykes et al.\ 1998	& \\
CLASS   & B2045+265 & 4	    & Fassnacht et al.\ 1999	& shear \\
\hline
JVAS    & B0218+357 & 2     & Patnaik et al.\ 1993  	& spiral \\
CLASS   & B0739+366 & 2	    & Marlow et al.\ 2001	& \\
JVAS    & B1030+074 & 2	    & Xanthopoulos et al.\ 1998	& faint satellite\\
CLASS   & B1127+385 & 2	    & Koopmans et al.\ 1999	& compound\\
CLASS   & B1152+199 & 2	    & Myers et al.\ 1999	& faint satellite\\
CLASS   & B1600+434 & 2	    & Jackson et al.\ 1995	& spiral, shear\\
CLASS   & B2319+051 & 2     & Rusin et al.\ 2001a       & shear\\
CLASS   & New lens  & 2     &                           & no optical imaging\\

\hline
CLASS   & B1359+154 & 6	    & Rusin et al.\ 2001b	& compound, shear\\
JVAS    & B1938+666 & 4+2+ring  & King et al.\ 1997     & \\
JVAS    & B2114+022 & 2? 4? & Augusto et al.\ 2001      & compound\\

\hline
\end{tabular}
\caption{Lens systems in the JVAS/CLASS sample. }
\end{table*}

\clearpage

\begin{figure*}
\begin{tabular}{c}
\psfig{file=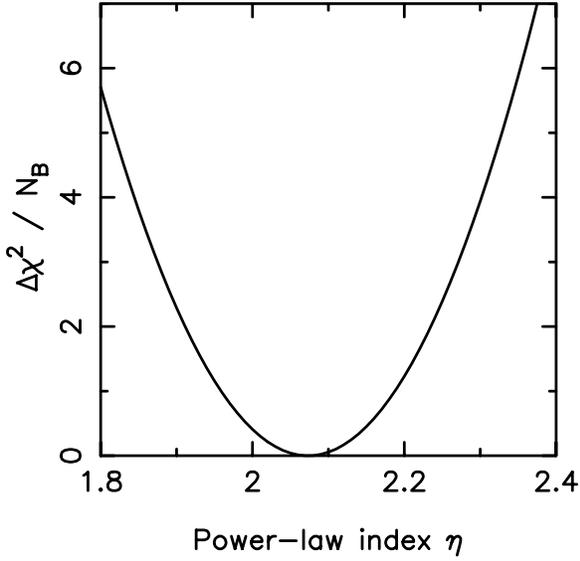,width=3in}\\
\end{tabular}
\figurenum{1}
\caption{Fit to the flat-spectrum number-flux relation for sources with flux
densities of $30 < S_5 < 200$ mJy at 5 GHz. The reduced $\chi^2$ is plotted
versus power-law $\eta$, where $dn / dS \propto S^{-\eta}$. The sources are
best fit by $\eta = 2.07$ ($\chi^2/N_B = 2.3$)}
\end{figure*}

\begin{figure*}
\psfig{file=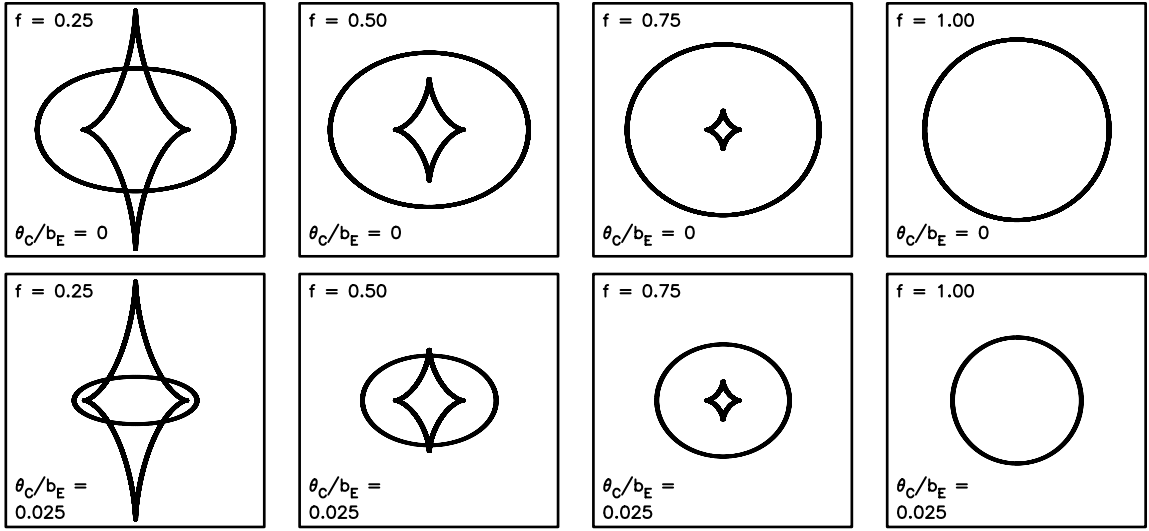,width=6in}
\figurenum{2}
\caption{Caustic plots for isothermal mass distributions. Top: singular
case. Bottom: $\theta_c / b_E = 0.025$. Left to right: $f$ = 0.25, 0.50, 0.75,
1.0.  Doubles are produced by sources within the radial (elliptical) caustic
only. Quads are produced by sources within both the radial and tangential
(astroid) caustics. Naked-cusp configurations are produced by sources within
the tangential caustic only. Note the increasing size of the tangential
caustic for flatter models, and the decreasing size of the radial caustic for
larger cores.}
\end{figure*}

\clearpage

\begin{figure*}
\begin{tabular}{c}
\psfig{file=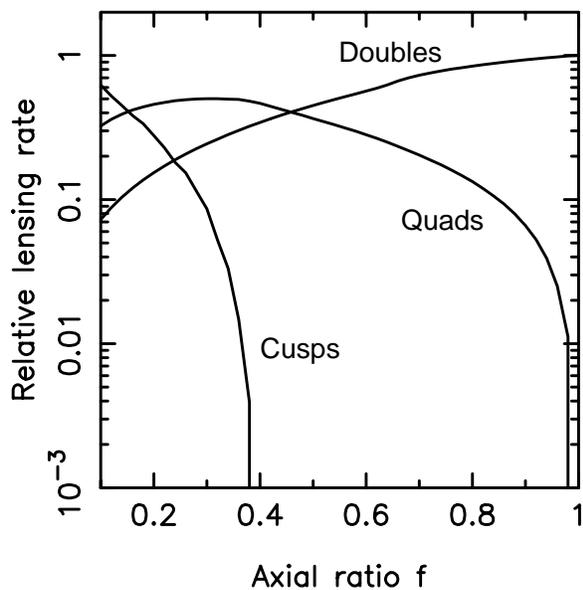,width=3in}
\end{tabular}
\figurenum{3}
\caption{Relative lensing rates for isothermal deflectors as a function of
axial ratio $f$. The radio luminosity function and survey selection effects
have been included.}
\end{figure*}

\begin{figure*}
\begin{tabular}{c}
\psfig{file=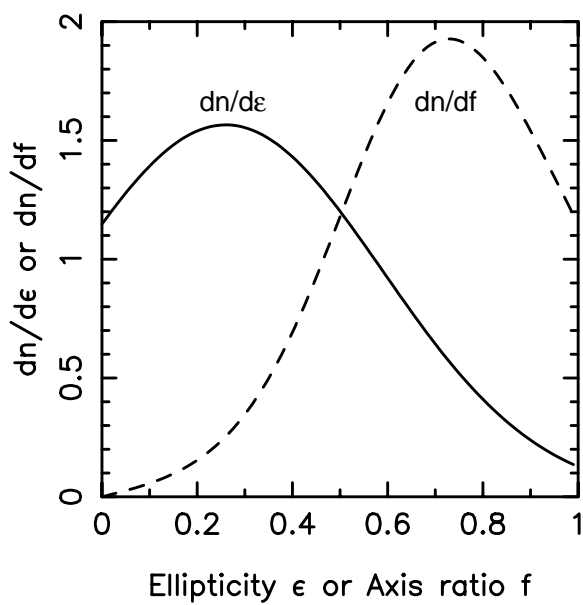,width=3in} \\
\end{tabular}
\figurenum{4}
\caption{The Coma ellipticity model. Plotted are the density of deflectors as
a function of ellipticity ($dn/d\epsilon$) and axial ratio ($dn/df$).}
\end{figure*}

\clearpage

\begin{figure*}
\begin{tabular}{c}
\psfig{file=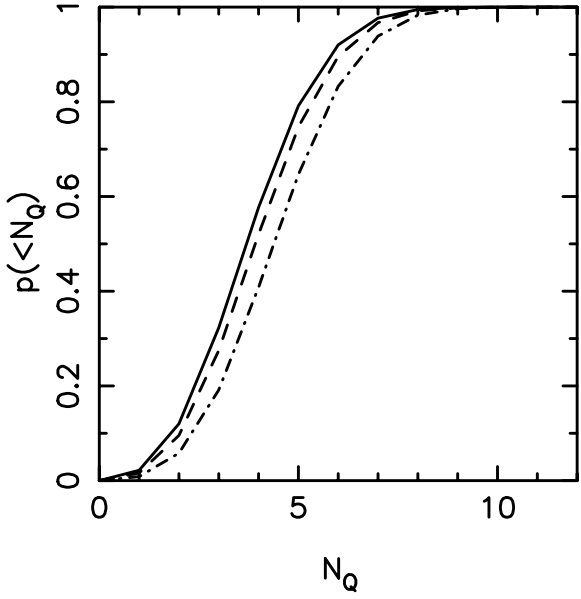,width=3in} \\
\end{tabular}
\figurenum{5}
\caption{Probability of the Coma model producing a sample of $N=12$ lenses
with fewer than $N_Q$ quads. Shown are the SIE model (solid), and SIE + shear
models with $\gamma = 0.05$ (dashed) and $\gamma = 0.10$ (dash-dotted). In 
the observed case of $N_Q = 7$, $p(<N_Q) = 97.7\%$, $96.7\%$ and $93.9\%$ for
the three models, respectively.}
\end{figure*}

\begin{figure*}
\begin{tabular}{c}
\psfig{file=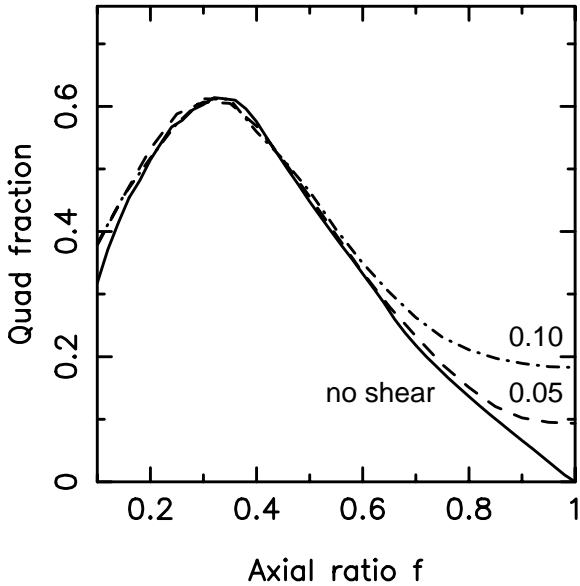,width=3in} \\
\end{tabular}
\figurenum{6}
\caption{The effect of external shear. Fraction of four-image lens systems as
a function of axial ratio $f$ for orientation-averaged shear fields of $\gamma
= 0$ (solid), $\gamma = 0.05$ (dashed) and $\gamma = 0.10$ (dash-dotted). Note
that the introduction of shear increases the quad fraction for $f \ga 0.7$.}
\end{figure*}

\clearpage

\begin{figure*}
\begin{tabular}{c c}
\psfig{file=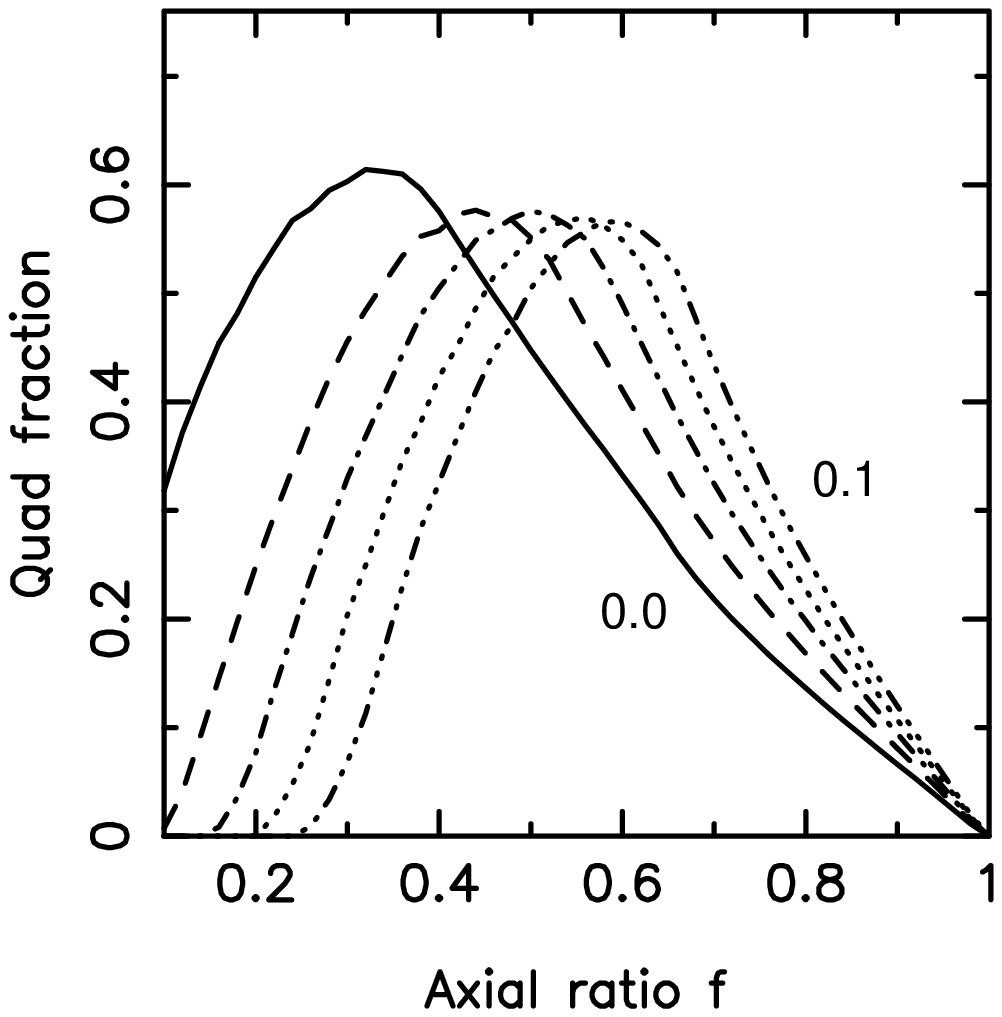,width=3in}&
\psfig{file=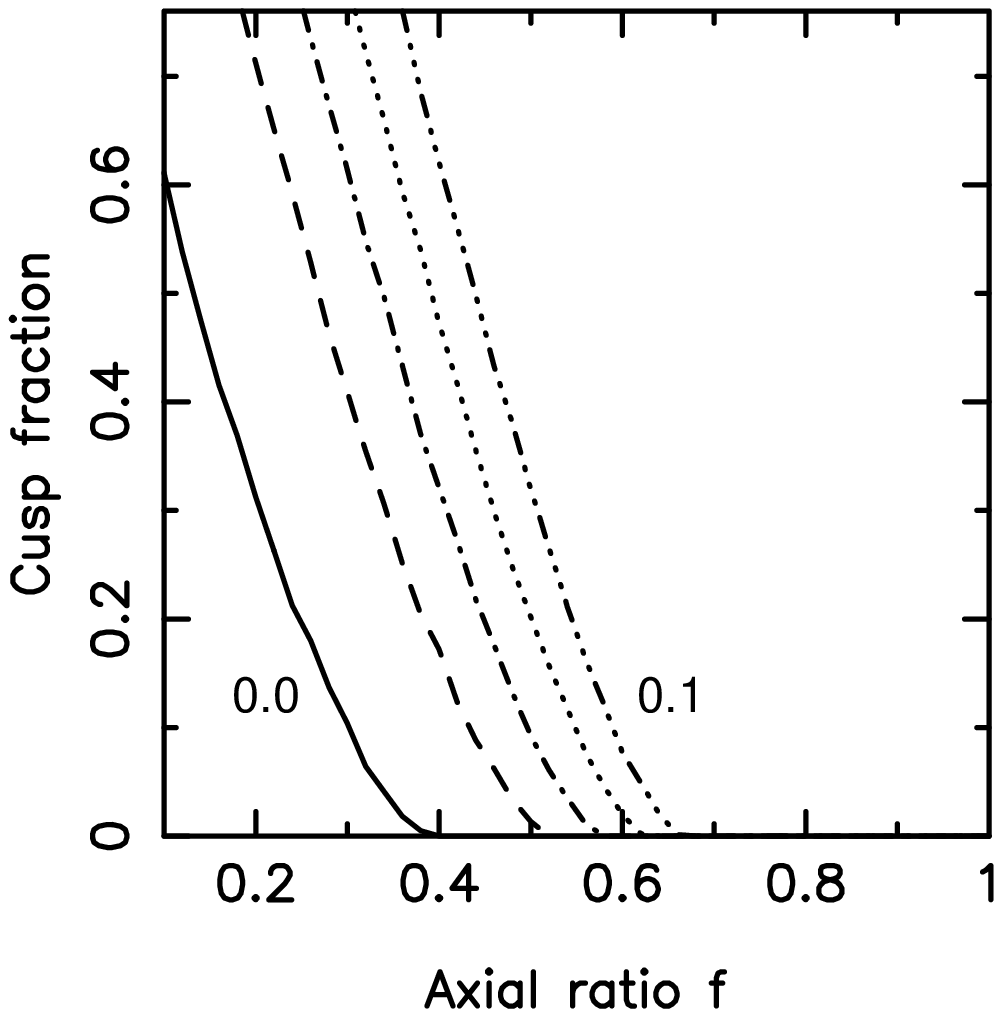,width=3in}\\
\psfig{file=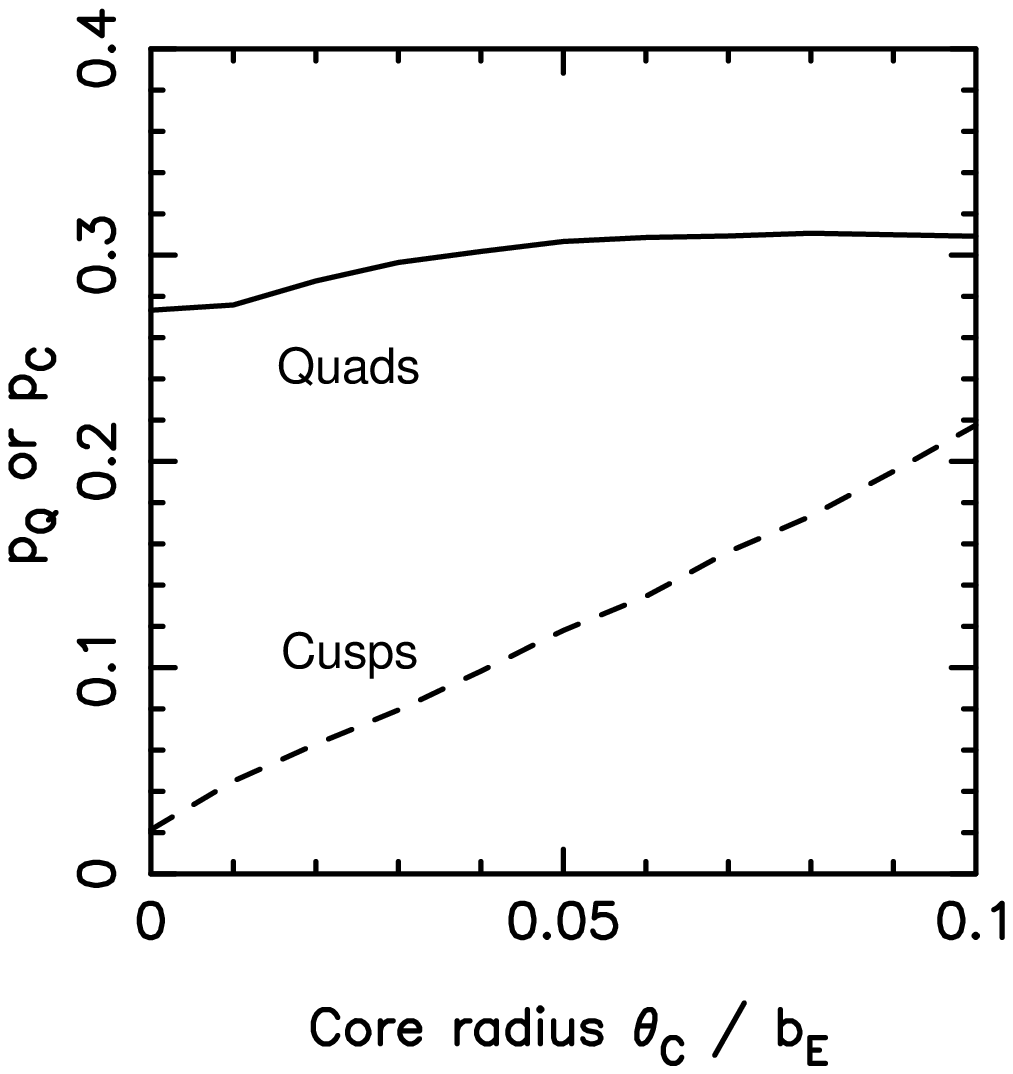,width=3in}&
\psfig{file=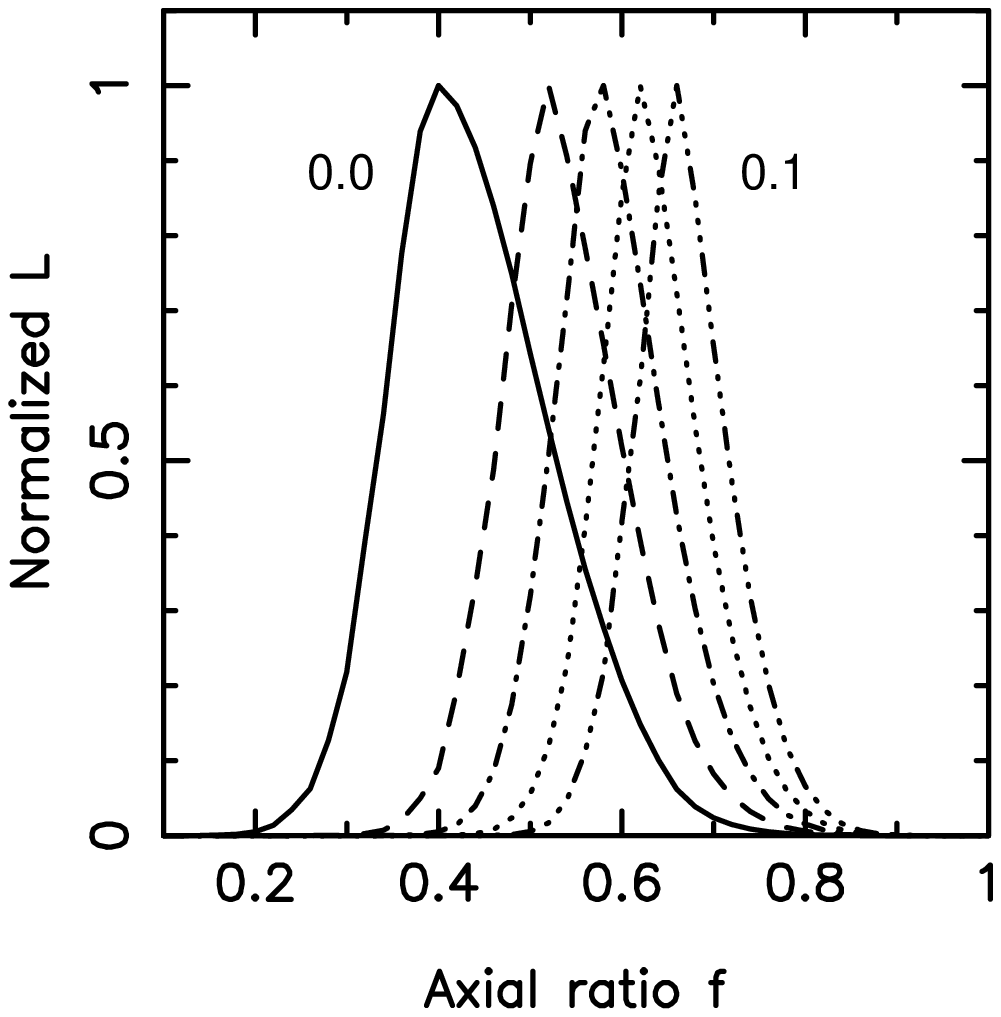,width=3in}\\
\end{tabular}
\figurenum{7}
\caption{The effect of cores in isothermal deflectors. (a) Top left: Quad
fraction for cores of size $\theta_c / b_E = 0$, 0.025, 0.050, 0.075, and
0.100. Note that cores increase the quad fraction for larger $f$, but decrease
it for smaller $f$. (b) Top right: naked-cusp fraction. Note that cores allow
for the production of these systems at larger axial ratios. (c) Bottom left:
Integrated fraction of quads ($p_Q$, solid) and cusps ($p_C$, dashed) for the
Coma ellipticity model, as a function of $\theta_c / b_E$. Note that there is
little change in the quad fraction as the core is increased, when integrating
over all ellipticities. (d) Bottom right: Preferred NIE axis ratio for five
different characteristic core radii, using (5).}
\end{figure*}

\begin{figure*}
\begin{tabular}{c c}
\psfig{file=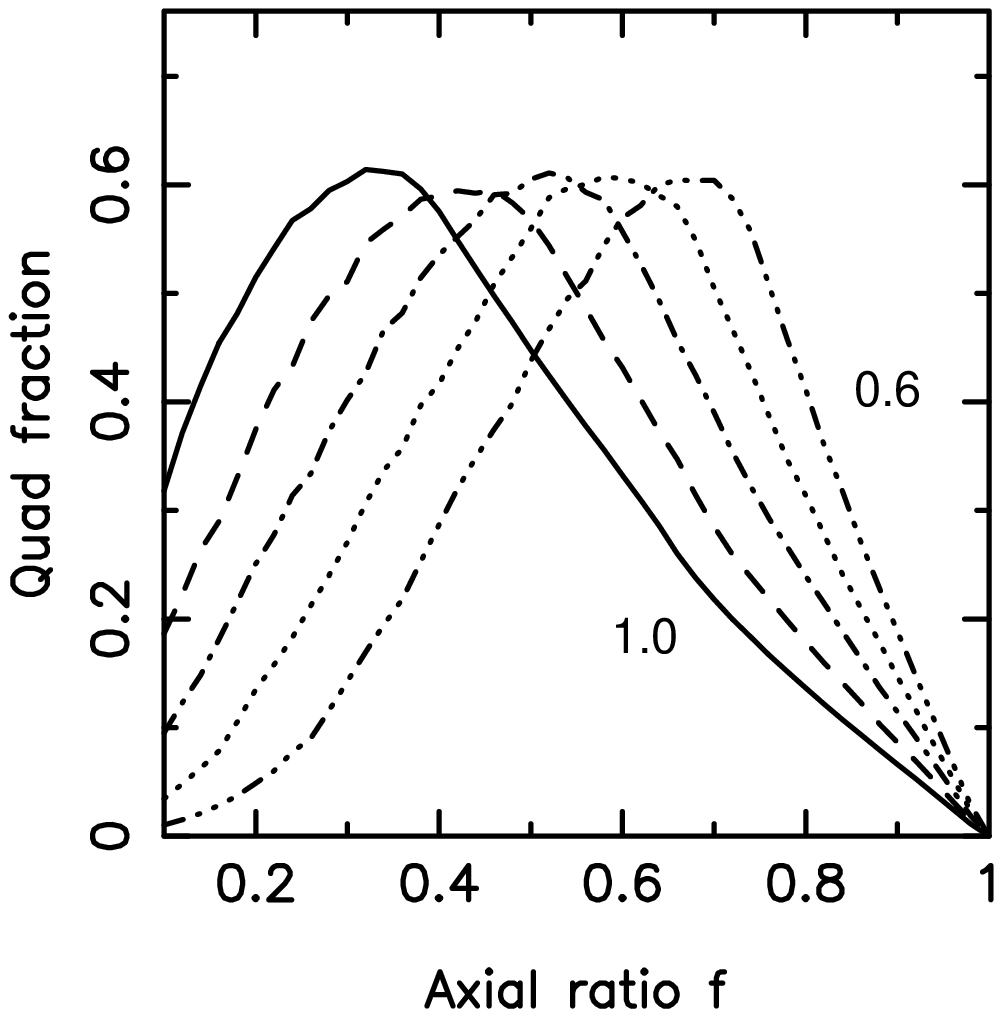,width=3in}&
\psfig{file=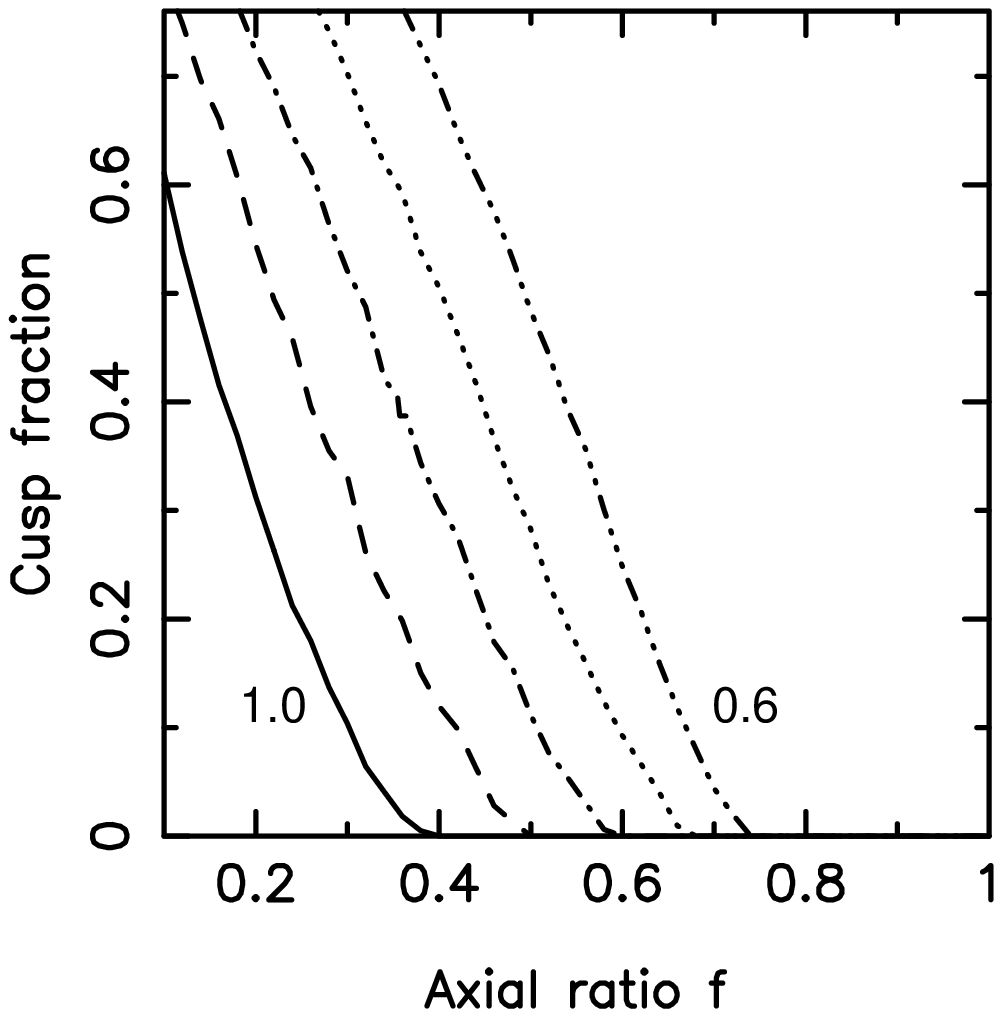,width=3in}\\
\psfig{file=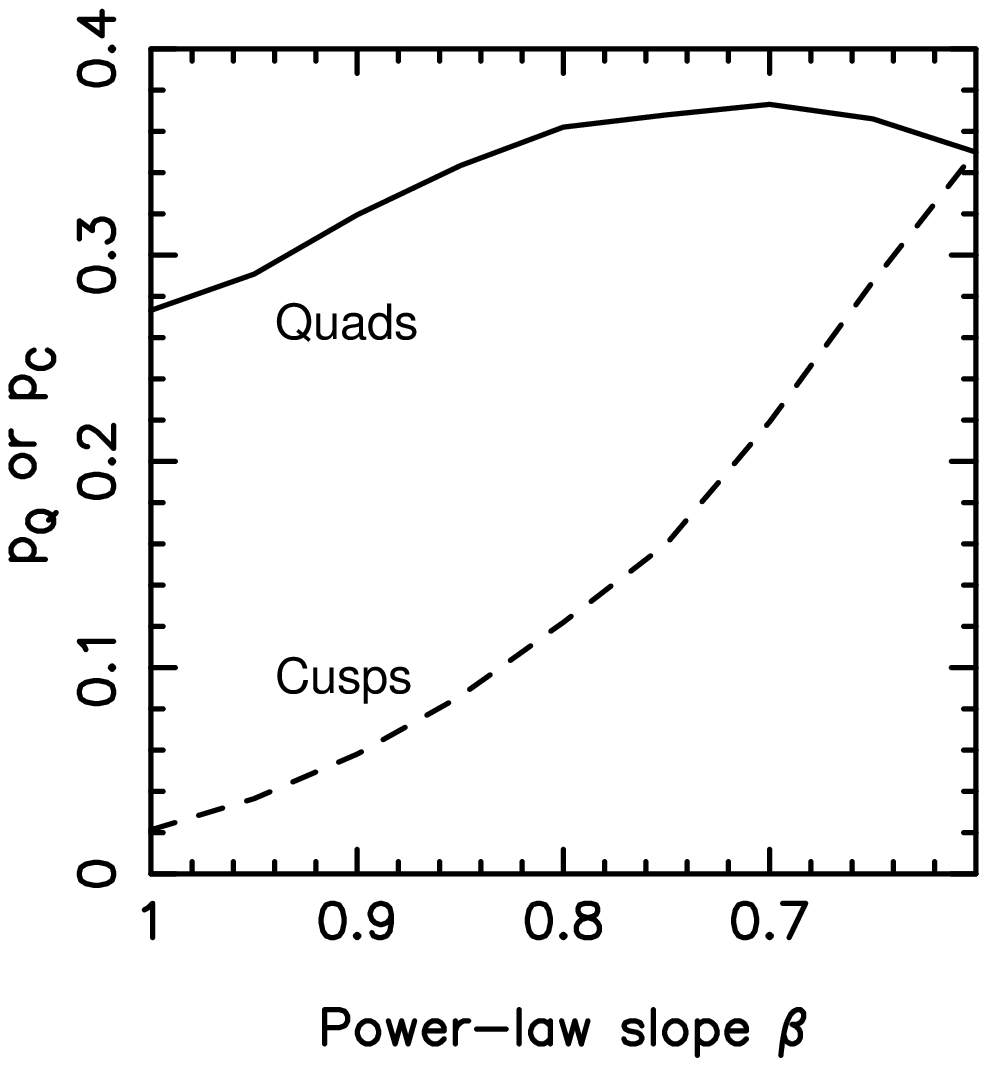,width=3in}&
\psfig{file=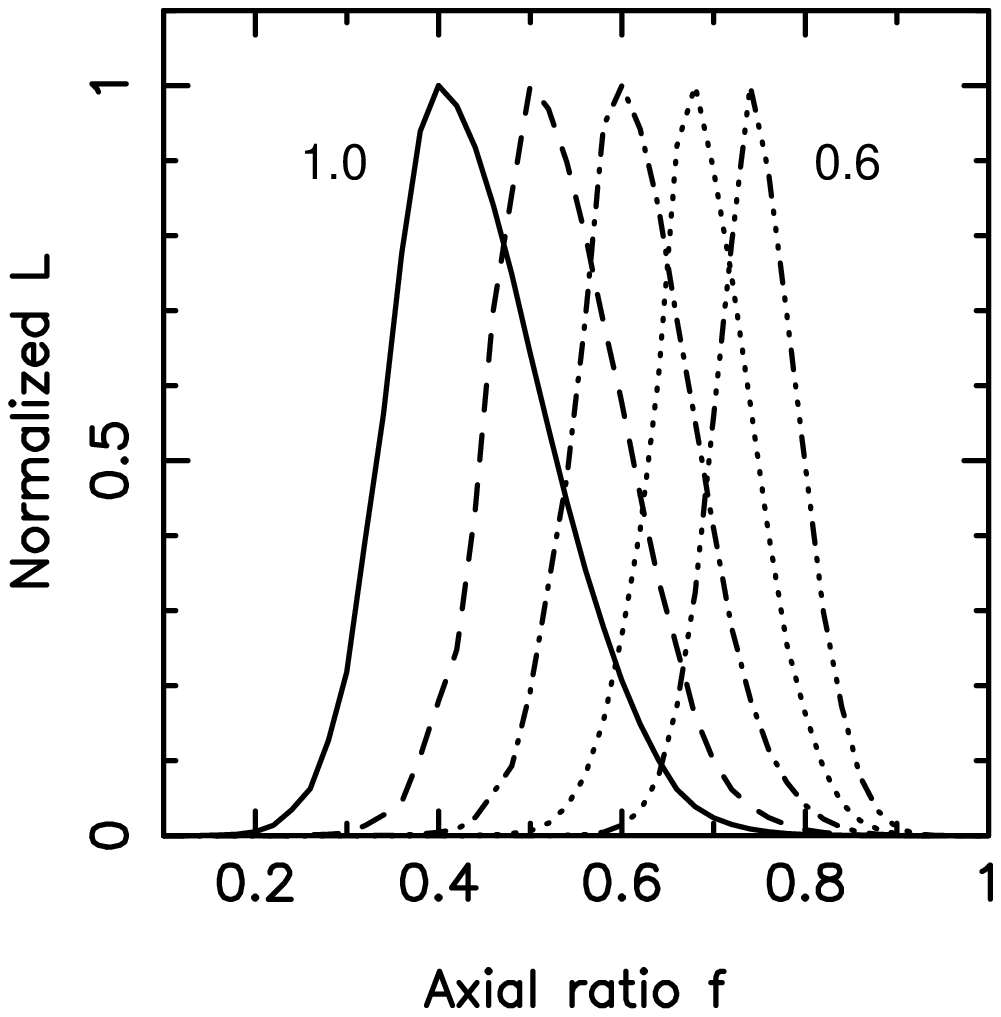,width=3in}\\
\end{tabular}
\figurenum{8}
\caption{The effect of shallow mass profiles. Same as in Fig.\ 7, but for 
profile slopes of $\beta = 1.0$, 0.9, 0.8, 0.7 and 0.6.} 
\end{figure*}

\begin{figure*}
\begin{tabular}{cc}
\psfig{file=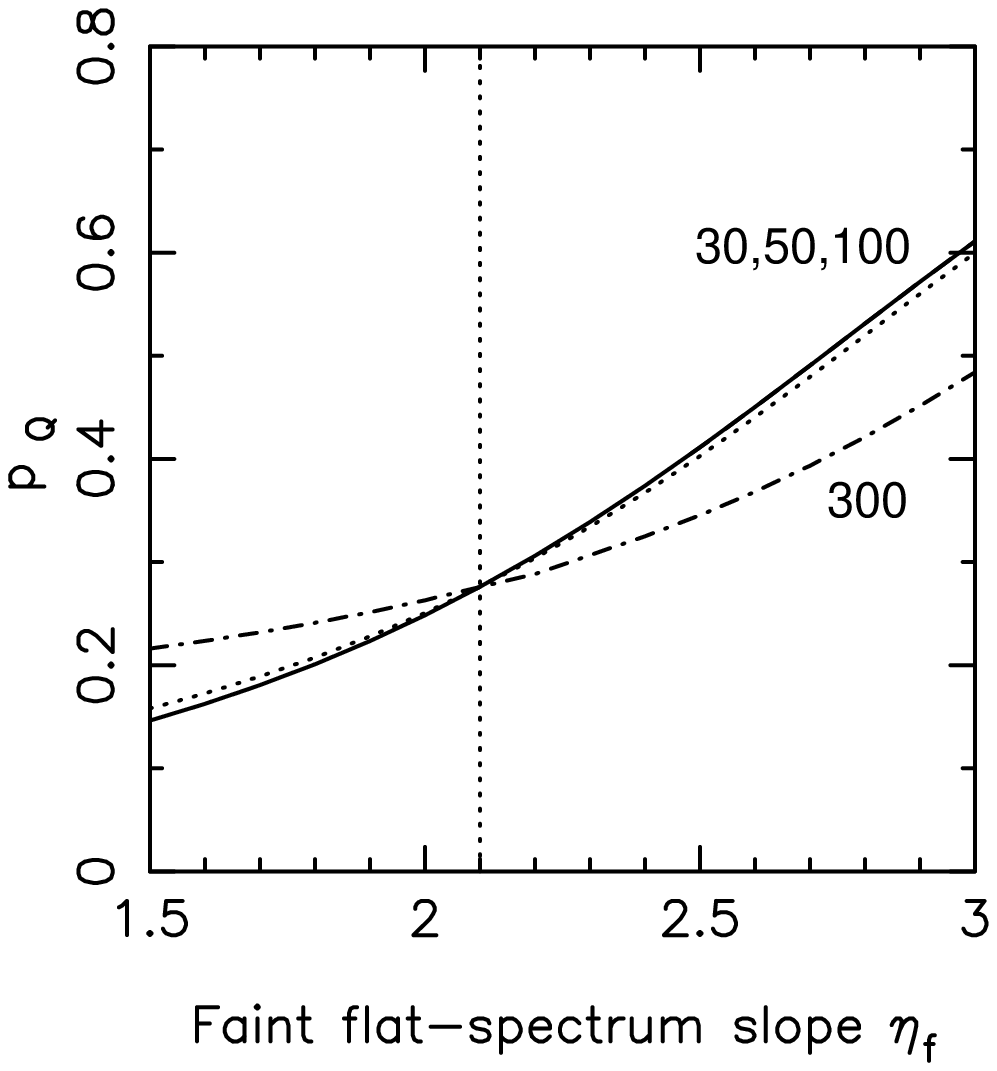,width=3.2in}&
\psfig{file=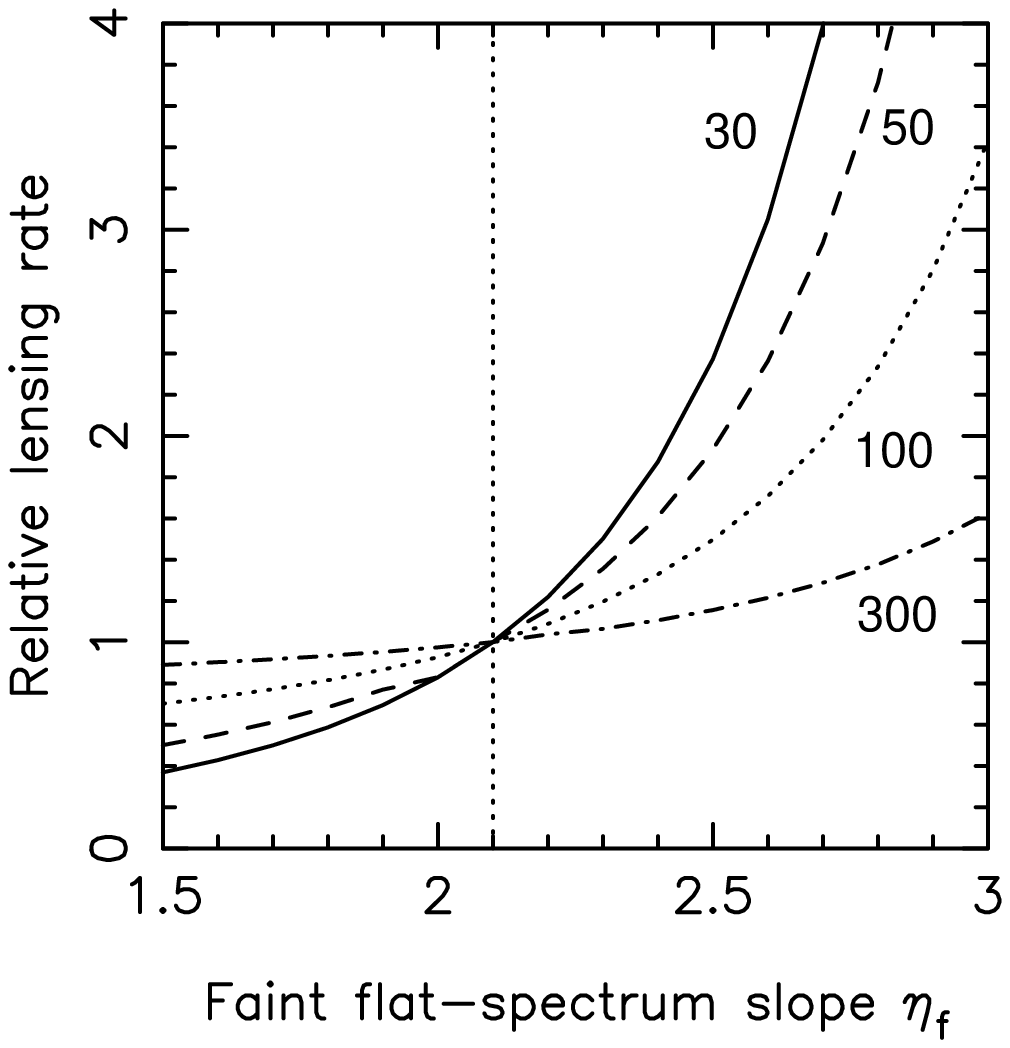,width=3.2in}\\
\end{tabular}
\figurenum{9}
\caption{The effect of a hypothetical break in the faint flat-spectrum
luminosity function at 30 mJy. (a) Left: Quad fraction. (b) Right: Total
lensing rate. Sources with flux densities of $S_5$ = 30, 50, 100 and 300 mJy
are plotted. The vertical marker denotes our initial assumptions of an
unbroken luminosity function. }
\end{figure*}

\end{document}